\documentclass[aip,jcp,reprint,amsmath,amssymb]{revtex4-2}
\usepackage{graphicx}
\usepackage[suffix=]{epstopdf}
\usepackage{natmove}

\newcommand{\vv}[1]{\mathbf{#1}}
\renewcommand{\d}[1]{\ensuremath{\operatorname{d}\!{#1}}}

\begin{document}

\title{Mesoscale simulations of diffusion and sedimentation in shape-anisotropic nanoparticle suspensions}

\author{Yashraj M. Wani}
\thanks{These authors contributed equally.}
\affiliation{Institute of Physics, Johannes Gutenberg University Mainz, Staudingerweg 7, 55128 Mainz, Germany}

\author{Penelope Grace Kovakas}
\thanks{These authors contributed equally.}
\affiliation{Department of Chemical Engineering, Auburn University, Auburn, Alabama 36849, USA}

\author{Arash Nikoubashman}
\email{anikouba@ipfdd.de}
\affiliation{Leibniz-Institut f{\"u}r Polymerforschung Dresden e.V., Hohe Stra{\ss}e 6, 01069 Dresden, Germany}
\affiliation{Institut f{\"u}r Theoretische Physik, Technische Universit{\"a}t Dresden, 01069 Dresden, Germany}

\author{Michael P. Howard}
\email{mphoward@auburn.edu}
\affiliation{Department of Chemical Engineering, Auburn University, Auburn, Alabama 36849, USA}

\begin{abstract}
We determine the long-time self-diffusion coefficient and sedimentation coefficient for suspensions of nanoparticles with anisotropic shapes (octahedra, cubes, tetrahedra, and spherocylinders) as a function of nanoparticle concentration using mesoscale simulations. We use a discrete particle model for the nanoparticles, and we account for solvent-mediated hydrodynamic interactions between nanoparticles using the multiparticle collision dynamics method. Our simulations are compared to theoretical predictions and experimental data from existing literature, demonstrating good agreement in the majority of cases. Further, we find that the self-diffusion coefficient of the regular polyhedral shapes can be estimated from that of a sphere whose diameter is average of their inscribed and circumscribed sphere diameters.
\end{abstract}
\maketitle

\section{Introduction}
The dynamics of nanoparticles (NPs) in suspensions play an important role in numerous applications, ranging from cellular transport \cite{sear:prl:2019} to the fabrication of functional nanomaterials \cite{boles:cr:2016}. For example, therapeutic agents can be encapsulated inside or attached to NPs for targeted drug delivery, and differences in NP dynamics in the body can affect their uptake and efficacy.\cite{Ding:mse:2017, Wong:cr:2015, Toy:nm:2014} Many factors impact the motion of NPs, including their size, interactions with each other, and interactions with their surroundings.\cite{wang:spus:2002} This work focuses specifically on the effect of shape, which has emerged as an important factor for modulating the properties and function of NPs in many practical applications \cite{yang:csr:2019, Andrade:ijms:2020}. For example, shape-anisotropic iron-oxide-based magnetic NPs were shown to enhance contrast for magnetic resonance imaging compared to spherical NPs,\cite{Andrade:ijms:2020} while quantum rods were shown to have enhanced diffusion compared to quantum dots in confined networks.\cite{rose:mm:2022} Given that NPs with a variety of shapes can be readily synthesized \cite{Yugang:sci:2002, Gou:nl:2003, Greyson:sm:2006} and that many naturally occurring NPs (e.g., the rod-like tobacco mosaic virus\cite{bawden:nat:1936, beijerinck:vka:1898} and gibbsite platelets\cite{kooij:nat:2000}) also exhibit pronounced shape anisotropy, it is important to develop a fundamental understanding of the relationship between shape and transport properties, such as diffusion coefficients, in order to engineer NPs for practical applications.

Experimentally characterizing how NP dynamics depend on shape and concentration can be challenging. For example, dynamic light scattering is a common technique for measuring NP diffusion from fluctuations in scattered light intensity\cite{Berne:2000}. However, knowledge about the distribution of NP sizes and/or shapes is needed to extract the diffusion coefficient from the raw measurement data \cite{Schartl:2007, hassan:lang:2015}, and it is difficult to perform this analysis for non-dilute solutions\cite{stetefeld:br:2016}. Camera-based tracking of tagged NPs is an alternative approach that allows for the direct measurement of the NP diffusion coefficient\cite{Lettinga:euLett:2005, rose:mm:2022}, but this method has limited spatial and temporal resolution\cite{Lettinga:euLett:2005}. NP properties may also be affected if labeling agents, such as fluorescent markers, are used \cite{Kundukad:sm:2014}. Further, it can be difficult to prepare NP suspensions with sufficiently low polydispersity and at high enough concentrations to accurately assess how the diffusion coefficient varies with both NP characteristics and concentration.

As a result, theory and simulations have proven to be useful approaches for studying the dynamics of NP suspensions. Early theories predominantly focused on spherical NPs, for which the single-particle translational and rotational diffusion coefficients can be calculated using the classical Stokes--Einstein and Stokes--Einstein--Debye relations, respectively. Theoretical predictions for the first-order concentration dependence of the long-time self-diffusion coefficient for suspensions of spherical NPs have also been derived.\cite{Brady:jfm:1994, tokuyama:pre:1994} Beyond spherical NPs, pioneering works by Kuhn, Kirkwood, and others have led to estimates for the single-particle translational and rotational diffusion coefficients of rod-like particles \cite{Kuhn:zpc:1932, kuhn:kz:1933, kuhn:hca:1945, riseman:jcp:1950, kuhn:kz:1933, tirado:jcp:1979, tirado:jcp:1980}. At finite concentration, the diffusive motion of the rods becomes more complex but can be split qualitatively into three regimes: at dilute concentrations, rods have essentially unrestricted motion in all directions; at semi-dilute concentrations, their motion is slightly hindered perpendicular to the long axis of the rod; and at high concentrations, the perpendicular diffusive motion is entirely suppressed.\cite{bruggen:pre:1998} However, predicting the dynamics of rod-like NPs with quantitative accuracy still remains challenging because their anisotropic shape can lead to complex flow patterns around individual NPs and to non-trivial collective behavior such as nematic or smectic ordering. For more complicated NP shapes than rods, predicting even single-particle diffusion coefficients becomes challenging, and numerical approaches are often required \cite{youngren:jfm:1975, okada:mp:2020}. In general, fully analytic descriptions of NP dynamics in suspensions are challenging to construct due to the many-body hydrodynamic interactions (HIs) between NPs that are mediated by the solvent.

Computer simulations are highly useful tools for numerically investigating NP dynamics in suspensions. The main challenge is to construct models that capture the relevant physics while remaining computationally tractable. Explicitly resolving both the NPs and the solvent molecules they are suspended in using, e.g., classical molecular dynamics (MD) approaches, quickly becomes infeasible because NPs are typically much larger than solvent molecules. However, given the corresponding separation of time scales between the solvent dynamics and NP dynamics, it is often possible to overcome this difficulty using coarse-grained models having simplified or implicit treatments of the solvent \cite{howard:coce:2019}. For example, Brownian dynamics (BD) is a well-known implicit-solvent technique that accounts for solvent-induced drag and fluctuating forces on the NPs\cite{Allen:oxford:2017}, but which neglects HIs between the NPs in its most basic form. HIs can be introduced to BD through appropriate mobility tensors\cite{Ermak:jcp:1978}, such as the pairwise far-field Rotne--Prager--Yamakawa tensor for spherical particles \cite{Rotne:jcp:1969, yamakawa:jcp:1970}. Stokesian dynamics, a gold-standard approach for simulating colloidal suspensions, additionally accounts for short-range lubrication forces between NPs within the BD framework \cite{brady:jfm:1988, brady:arfm:1988}. However, BD approaches that include HIs are often still computationally demanding to implement and require expressions for the mobility tensor, which may be difficult to obtain for complex NP shapes.

To circumvent issues determining inputs needed for a fully implicit treatment of the solvent, several mesoscale simulation methods, including multiparticle collision dynamics (MPCD)\cite{howard:coce:2019, malevanets:jcp:1999, Gompper:springer:2009}, dissipative particle dynamics \cite{hoogerbrugge:eurolett:1992, Liu:acme:2015}, and the lattice Boltzmann method \cite{Dunweg:springer:2009, Ladd:jsp:2001}, use simplified particle-based solvent models that are less demanding to simulate than an atomistic model but still have properties resembling that of real solvents. In this work, we will use MPCD because we have recently shown that MPCD can reasonably reproduce expected results for the long-time self-diffusion coefficient and sedimentation coefficient for suspensions of spherical NPs over a range of NP concentrations \cite{wani:jcp:2022}, and the same approach used to model the spherical NPs can be extended to NPs with other shapes. In MPCD, NPs are modeled as conventional MD particles that can be coupled to the solvent through different schemes to ensure HIs develop \cite{poblete:pre:2014, wani:jcp:2022, malevanets:jcp:2000, nikoubashman:sm:2013, batot:pre:2013, dahirel:pre:2018, padding:pre:2006}. The current state-of-the-art coupling scheme, first proposed by Poblete et al., uses a discrete particle model that represents an NP as a mesh of ``vertex'' particles interconnected via elastic springs \cite{poblete:pre:2014}. The solvent particles interact with the NPs only through stochastic collisions that are straightforward to compute. We used a discrete particle model to study the long-time self-diffusion of cubes,\cite{wani:jcp:2022} and similar models have been used to simulate the self-assembly of colloids with shape and/or interaction anisotropy \cite{kobayashi:lang:2020, kobayashi:sm:2020, kobayashi:lang:2020, kobayashi:sm:2020, yokoyama:sm:2023, kobayashi:lng:2022, argun:sm:2023, ikeda:msde:2024}. However, we are unaware of a systematic study using MPCD to characterize the long-time self-diffusion coefficients and sedimentation coefficients for suspensions of shape-anisotropic NPs at varying concentrations.

In this work, we use MPCD with a discrete particle model to study the long-time self-diffusion and sedimentation coefficients of octahedra, cubes, tetrahedra, and spherocylinders as a function of NP concentration. We investigate the effect of shape by comparing the results for the different NP shapes with each other and with spheres. We also assess the influence of solvent-mediated HIs by comparing the MPCD simulations with implicit-solvent Langevin dynamics simulations.

\section{Models}
\subsection{Multiparticle collision dynamics}
In MPCD, the solvent consists of point particles that are propagated in alternating streaming and collision steps that occur at a regular time interval $\Delta t$. During the streaming step, the solvent particles move according to Newton's equations of motion,
\begin{align} \label{eq:newton}
\frac{\d{\vv{r}_i}}{\d{t}} &= \vv{v}_i \nonumber \\
m_i \frac{\d{\vv{v}_i}}{\d{t}} &= \vv{F}_i,
\end{align}
where $\vv{r}_i$ is the position, $\vv{v}_i$ is the velocity, and $m_i$ is the mass of particle $i$. All solvent particles have the same mass $m$. Unlike standard MD particles, MPCD particles do not interact with each other by pairwise forces, but each particle may be acted on by a body force $\vv{F}_i$. For a constant $\vv{F}_i$, eq.~\eqref{eq:newton} can be integrated analytically to give the standard equations of ballistic motion.

In the collision step, the solvent particles are sorted into cubic cells of edge length $\ell$, then exchange momentum with particles in the same cell according to a collision scheme. Here, we use the stochastic rotation dynamics (SRD) scheme without angular momentum conservation \cite{malevanets:jcp:1999, Gompper:springer:2009}. SRD updates the velocity of particle $i$ in cell $j$ according to:
\begin{equation} \label{collision}
\vv{v}_i \gets \vv{u}_j + \boldsymbol{\Omega}_j \cdot (\vv{v}_i - \vv{u}_j),
\end{equation}
where $\vv{u}_j$ is the mass-averaged velocity of the particles in cell $j$ and $\boldsymbol{\Omega}_j$ is the rotation matrix for cell $j$. The matrix $\boldsymbol{\Omega}_j$ rotates about an axis randomly selected for cell $j$ by a fixed angle $\alpha$. At each collision step, the collision cells are shifted along each Cartesian direction by a random amount drawn uniformly from $[-\ell/2, +\ell/2]$ to ensure Galilean invariance, \cite{ilhe:pre:2001, ilhe:pre:2003} and a cell-level Maxwellian thermostat is used to maintain a constant temperature $T$ \cite{huang:jcp:2010}.

The natural units for MPCD simulations are the length $\ell$ of the collision cells, the mass $m$ of the solvent particles, and the thermal energy $k_{\rm B}T$, where $k_{\rm B}$ is the Boltzmann constant. The corresponding unit of time is $\tau = \sqrt{m \ell^2 \beta}$, where $\beta = 1/(k_{\rm B} T)$. We adopted the standard SRD parameters $\Delta t = 0.1\,\tau$, $\alpha = 130^\circ$, and average solvent number density $5\,\ell^{-3}$, which give a liquid-like Newtonian fluid with dynamic viscosity $\eta_0 = 3.95\,k_{\rm B}T\tau/\ell^3$\cite{statt:prf:2019}.

\subsection{Discrete particle model}
\begin{figure*}
  \includegraphics{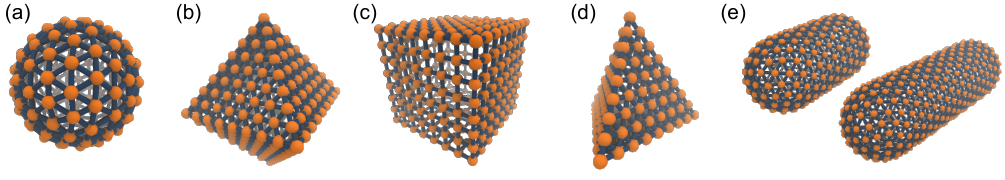}
  \caption{Discrete particle model for (a) sphere, (b) octahedron, (c) cube, (d) tetrahedron and (e) two spherocylinders (aspect ratios $\lambda = 1$ and 2). The $N_{\rm v}$ vertex particles are shown in orange, and the bonds to their nearest neighbors are shown in blue. To improve the clarity of these renderings, the size of the vertex particles has been decreased, and central particles and additional bonds used to maintain the shape have been omitted. These snapshots were rendered using Visual Molecular Dynamics (version 1.9.3)\cite{humphrey:jmg:1996}}
  \label{fig:particles}
\end{figure*}

A discrete particle model was used to represent the NPs and couple them to the solvent \cite{poblete:pre:2014, wani:jcp:2022}. The NP shapes we modeled were a sphere, an octahedron, a cube, a tetrahedron, and two spherocylinders with different aspect ratios (Fig.~\ref{fig:particles}). Each NP consisted of $N_\text{v}$ vertex particles on the surface of the shape, and each vertex particle had mass $5\,m$. The vertex particles were bonded to their nearest neighbors with a harmonic potential,
\begin{equation}
\beta u_{\rm b}(r) = \frac{k_{\rm b}}{2}(r-r_{\rm b})^2 ,
\label{eq:harmonic}
\end{equation}
where $r$ is the distance between two particles, $r_{\rm b}$ is the distance required for the bond by the shape, and $k_{\rm b}$ is the spring constant. To ensure that the NPs maintained their shapes, the vertex particles were also bonded to either an additional particle in the center of the NP (sphere, octahedron, cube, \& tetrahedron) or their diametrically opposed vertex particle (spherocylinders). Excluded-volume interactions between NPs were modeled by applying the Weeks--Chandler--Andersen repulsive potential\cite{weeks:jcp:1971} between vertex particles
\begin{equation}
\beta u(r) = \begin{cases}
    \displaystyle 4 \left[ \left(\frac{\sigma}{r}\right)^{12} - \left(\frac{\sigma}{r}\right)^{6} \right] + 1,  & r \le 2^{1/6}\,\sigma
    \\
    0, &\text{otherwise}
\end{cases}.
\label{eq:wca}
\end{equation}
All vertex particles (but not the central particle) were coupled to the MPCD solvent through the collision step eq.~\eqref{collision} \cite{poblete:pre:2014}. Between collision steps, the central and vertex particles moved according to eq.~\eqref{eq:newton}. Based on our prior work \cite{wani:jcp:2022}, we used $k_{\rm b} = 5000\,\ell^{-2}$ to make stiff bonds and $\sigma = \ell$, and we integrated eq.~\eqref{eq:newton} using the velocity Verlet algorithm with time step $0.005\,\tau$. We visually confirmed that all NPs maintained a nearly rigid shape and that no NPs penetrated each other for the vertex-particle configurations chosen as described next.

\textit{Sphere.}---We modeled a sphere having diameter $d = 6\,\ell$ [Fig.~\ref{fig:particles}(a)] as a reference point. To create the vertex particles, we subdivided the triangular faces of a regular icosahedron twice and scaled the positions of all vertices to lie on the surface of the sphere. This process resulted in $N_\text{v} = 162$ vertex particles with a typical nearest-neighbor distance between $0.83\,\ell$ and $0.97\,\ell$. Note that this model differs from the one we used in ref.~\citenum{wani:jcp:2022} in two ways: (1) the number of vertex particles is larger and (2) the excluded volume is handled through the vertex particles rather than through the central particle. These choices were made in this work so that the spheres would have a comparable surface density of vertex particles and the same style of excluded-volume interactions as the anisotropic NPs we studied.

\textit{Octahedron and tetrahedron.}---We modeled a regular octahedron and a regular tetrahedron both having edge length $a = 6\,\ell$ [Figs.~\ref{fig:particles}(b) and \ref{fig:particles}(d)]. Because the faces of these polyhedra are equilateral triangles, we first created a three-dimensional triangulated model of each shape using computer-aided design software, then subdivided the faces 3 times to create a triangular mesh of vertex particles. This process resulted in 9 vertex particles per edge and distance $a / 8 = 0.75\,\ell$ between all nearest-neighbor vertex particles for both shapes. The total number of vertex particles was $N_\text{v} = 258$ for the octahedron and $N_\text{v} = 130$ for the tetrahedron.

\textit{Cube.}---We modeled a cube with edge length $a = 6\,\ell$ [Fig.~\ref{fig:particles}(c)] using a square mesh with 8 vertex particles per edge. The total number of vertex particles was $N_\text{v} = 296$, and the distance between nearest-neighbor vertex particles was $a / 7 \approx 0.86\,\ell$. This is the same vertex particle configuration as in ref.~\citenum{wani:jcp:2022}, and the description here corrects a typographical error for the number of vertex particles per edge. Unlike in ref.~\citenum{wani:jcp:2022}, though, a central particle was used to maintain rigidity to keep consistency with the sphere, octahedron, and tetrahedron.

\textit{Spherocylinder.}---We modeled two types of spherocylinders: both had two hemispheres with diameter $d = 6\,\ell$, but one had a cylinder of length $h = 6\,\ell$ while the other cylinder had the length $h = 12\,\ell$ [Fig.~\ref{fig:particles}(e)]. Thus, the spherocylinders had aspect ratios $\lambda \equiv d/h = 1$ and 2, respectively. This degree of anisotropy is much smaller than that of rods typically used in experiments, which usually are in the range of $\lambda = 10$ to 150 \cite{lettinga:sm:2010}. However, we found surprisingly little data on transport coefficients for rod-like NPs with these smaller aspect ratios and, as noted previously, theories are also hard to formulate in this regime \cite{dhont:mm:1999}. Hence, we chose to study spherocylinders with these smaller aspect ratios to begin to bridge the knowledge gap between spheres and long rod-like NPs. Discrete particle models for the spherocylinders were constructed through a multi-step process: First, a mesh of vertex particles for the hemispheres was created by slicing our discrete sphere model in half along a plane that exposed 20 evenly spaced vertex particles around its circumference and had 91 vertex particles in total. Then, vertex particles for the cylinder were generated from the ring of 20 exposed vertex particles by translating the ring by $0.75 \,\ell$ and rotating it around the axis of the cylinder by $9^\circ$ to stagger the particles on consecutive rings. This process was repeated until the entire cylinder surface was covered with vertex particles. The total number of vertex particles per spherocylinder was $N_\text{v} = 322$ for $\lambda = 1$ and $N_\text{v} = 482$ for $\lambda = 2$, with the nearest-neighbor distance between vertex particles ranging from $0.83\,\ell$ to $0.97\,\ell$.

\subsection{Simulation details}
We performed bulk simulations containing $N$ NPs in a cubic simulation box with edge length $L=120\,\ell$ and periodic boundary conditions. We simulated a range of nominal NP volume fractions $\phi = Nv/L^3$, where $v$ is the nominal volume of each NP (Table~\ref{table:shape-specific-params}), by varying $N$. We created equilibrated configurations of NPs at the different volume fractions using Langevin dynamics (LD) simulations. LD simulations are faster to perform than MPCD simulations because they do not include HI, and we also chose the friction coefficient for the LD simulations to give faster NP dynamics than in the MPCD simulations in order to accelerate equilibration. Starting from these configurations, we measured the long-time self-diffusion coefficient as a function of $\phi$ using equilibrium simulations (Section \ref{sec:diffusion}) and the sedimentation coefficient as a function of $\phi$ using nonequilibrium simulations (Section \ref{sec:sediment}). All simulations were conducted using HOOMD-blue \cite{anderson:cms:2020, howard:cpc:2018} (version 2.9.7) extended with azplugins\cite{howard:github} (version 0.12.0).

\begin{table*}
\caption{Geometric properties of the regular polyhedra investigated. General formulae are given in terms of the edge length $a$, with the specific value for $a = 6\,\ell$ (the edge length for all our polyhedral NPs) quoted in parentheses. The properties are the volume $v$, surface area $A$, sphericity $\psi$, inscribed-sphere  diameter $d_{\rm I}$, circumscribed-sphere diameter $d_{\rm C}$, and mean of inscribed-sphere and circumscribed-sphere diameters $\bar d$.}
\label{table:shape-specific-params}
\begin{tabular}{c|cc|cc|c|cc|cc|cc}
     & $v$ & ($\ell^3$) & $A$ & ($\ell^2$) & $\psi$ & $d_{\rm I}$ & ($\ell$) & $d_{\rm C}$ & ($\ell$) & $\bar d$ & ($\ell$) \\ \hline
     octahedron & $\dfrac{\sqrt{2}}{3} a^3$ & (101.8) & $2\sqrt{3}a^2$ & (124.7) & 0.846 & $\sqrt{\dfrac{2}{3}}a$ & (4.9)& $\sqrt{2}a$ & (8.5) & $\dfrac{\sqrt{2}+\sqrt{3}}{\sqrt{6}} a$ & (6.7) \\[10pt]
     cube & $a^3$ & (216.0) & $6a^2$ & (216.0) & 0.806 & $a$ & (6.0) & $\sqrt{3}a$ & (10.4) & $\dfrac{1+\sqrt{3}}{2}a$ & (8.2) \\[10pt]
     tetrahedron & $\dfrac{a^3}{6\sqrt{2}}$ & (25.5) & $\sqrt{3} a^2$ & (62.4) & 0.671 & $\dfrac{a}{\sqrt{6}}$ & (2.4) & $\sqrt{\dfrac{3}{2}} a$ & (7.3) & $\sqrt{\dfrac{2}{3}}a$ & (4.9)
\end{tabular}
\end{table*}

For the spheres and regular polyhedra, we performed one equilibrium simulation of length $2 \times 10^5\,\tau$ and recorded the position of all central particles every $10\,\tau$. We performed one nonequilibrium simulation consisting of a warmup period of $0.5\times 10^5\,\tau$ to reach steady state followed by a production period of length $1.5 \times 10^5\,\tau$ in which we recorded the average velocity of the NPs every $0.105\,\tau$ and the average velocity of the solvent every $0.1\,\tau$. To estimate error bars, we subdivided these trajectories into three blocks and computed the standard error between blocks.

For the spherocylinders, we performed eight equilibrium simulations of length $10^5\,\tau$ and recorded the position of enough vertex particles to reconstruct the center of mass of each NP every $2.5\,\tau$. We performed three nonequilibrium simulations consisting of a $0.5 \times 10^5\,\tau$ warmup period and $1 \times 10^5\,\tau$ production period with the velocities sampled in the same way as for the other shapes. Error bars were estimated as the standard error of the independent simulations. 

\section{Results and Discussion}
\subsection{Long-time self-diffusion coefficient}
\label{sec:diffusion}
We computed the long-time self-diffusion coefficient $D$ of the NPs from the time derivative of the average mean squared displacement $\langle \Delta r^2 \rangle$ of each NP \cite{Allen:oxford:2017},
\begin{equation}
    D = \lim_{t \to \infty} \frac{1}{6} \frac{\d{\langle \Delta r^2 \rangle}}{\d{t}}.
\end{equation}
To improve statistics, we averaged $\langle \Delta r^2 \rangle$ over NPs and time origins, and we extracted $D$ from the time average of the long-time plateau of $\d{\langle \Delta r^2 \rangle}/\d{t}$, which we fit in the time range $10^4\,\tau \le t < 2\times 10^4\,\tau$ for the spheres and regular polyhedra and in the range $3 \times 10^4\,\tau \le t < 5 \times 10^4\,\tau$ for the spherocylinders. Note that in defining $D$ in this way, the long-time self-diffusion coefficient is a scalar quantity. For anisotropic NPs, the short-time motion is characterized by a diffusion tensor; this tensor is isotropic for the regular polyhedra we have studied \cite{Happel:sn:1983}, but it is anisotropic for the spherocylinders. \cite{doi:oup:1988} Hence, $D$ reported in this work implies an orientational average at long times for the spherocylinders.

Due to the long-ranged nature of solvent-mediated HIs, self-diffusion coefficients measured in simulations with periodic boundary conditions can suffer from noticeable finite-size effects \cite{dunweg:prl:1991, dunweg:jcp:1993, yeh:jpcb:2004}. For a cubic simulation box such as ours, the self-diffusion coefficient in an infinitely large box $D^\infty$ is related to $D$ in a finite box with edge length $L$ by \cite{yeh:jpcb:2004, dunweg:jcp:1993}
\begin{equation} \label{eq:D-FS-correct}
    D^\infty = D + \xi \frac{k_BT}{6\pi\eta L},
\end{equation}
where $\xi \approx 2.837297$ and $\eta$ is the suspension viscosity. Applying eq.~\eqref{eq:D-FS-correct} can be challenging in practice because it requires knowledge of $\eta$, which depends on the shape and volume fraction of the NPs. Analytic expressions for $\eta$ exist for some NP shapes,\cite{einstein:ap:1906, einstein:ap:1911} but they are typically only valid for small NP volume fractions\cite{Bolintineanu:cpm:2014}; hence, additional costly simulations are usually needed to accurately determine $\eta$. To avoid this step, we approximated $\eta$ with a Stokes--Einstein-like proportionality, $\eta/\eta_0 = D_0/D^\infty$ \cite{yeh:jpcb:2004, wani:jcp:2022}, where $D_0 = k_{\rm B} T/\gamma_0$ is the long-time self-diffusion coefficient at infinite dilution (i.e., the single-particle limit) and $\gamma_0$ is the corresponding hydrodynamic friction coefficient for the NP (again, orientationally averaged for the spherocylinders). Substituting for $\eta$ in eq.~\eqref{eq:D-FS-correct} and solving for $D^\infty$ yields
\begin{equation} \label{eq:D-FS-approx}
    D^\infty \approx D \left( 1 - \xi \frac{\gamma_0}{6 \pi \eta_0 L} \right)^{-1} .
\end{equation}
We previously tested this approximate correction by computing $D$ for spherical NPs in different box sizes $L$ and confirming that $D^\infty$ was independent of $L$ within our measurement accuracy.\cite{wani:jcp:2022}

To apply eq.~\eqref{eq:D-FS-approx}, $\gamma_0$ must be determined for each NP shape. Experimental correlations \cite{pettyjohn:cep:1948} for $\gamma_0$ exist [e.g., eq.~\eqref{eq:friction-theory} below]; however, it is not guaranteed that the MPCD simulations are consistent with these. Instead, we noted that all diffusivities are corrected by the same factor in eq.~\eqref{eq:D-FS-approx} regardless of $\phi$ and that eq.~\eqref{eq:D-FS-correct} can be used directly when $\phi$ is sufficiently small that $\eta \approx \eta_0$. Accordingly, we linearly extrapolated our measured $D$ to $\phi = 0$, using the data from the smallest two values of $\phi$ that we simulated, to obtain a measured $D_0$ with finite-size effects. We then applied eq.~\eqref{eq:D-FS-correct} with $\eta = \eta_0$ to calculate $D_0^\infty$ from $D_0$ and used the ratio $D_0^\infty / D_0$ as the finite-size correction factor for all $D$. In the rest of the paper, all diffusion coefficients have been corrected for finite-size effects in this way, but we will still refer to them as $D$ and $D_0$ for notational simplicity.

\subsubsection{Regular polyhedra}
We first investigated the shape-dependence of the long-time self-diffusion coefficient extrapolated to infinite dilution $D_0$ for the regular polyhedra we simulated (octahedron, cube, and tetrahedron). Pettyjohn and Christiansen experimentally measured the settling rates of particles with these shapes at low Reynolds number \cite{pettyjohn:cep:1948, Happel:sn:1983}. They found that the settling rate could be correlated with particle shape using the sphericity $\psi$, defined as the ratio of the surface area of a sphere having the same volume as the shape to the actual surface area $A$ of the shape,
\begin{equation} \label{eq:sphericity}
    \psi = \frac{\pi^{1/3} (6v)^{2/3}}{A}.
\end{equation}
The sphericities of our regular polyhedra are listed in Table \ref{table:shape-specific-params}. Using the correlation for the settling velocity from ref.~\citenum{pettyjohn:cep:1948}, a correlation for the hydrodynamic friction coefficient $\gamma_0$ is
\begin{equation} \label{eq:friction-theory}
\gamma_0 = 3 \pi \eta_0 \left( \frac{6v}{\pi} \right)^{1/3} \left[0.843 \log_{10} \left(\frac{\psi}{0.065}\right) \right]^{-1}.
\end{equation}
Note that the first term in parentheses is the diameter of an equivalent-volume sphere to the shape, so eq.~\eqref{eq:friction-theory} gives $\gamma_0 = 3\pi\eta_0 d$ for a sphere with diameter $d$ as expected.

Based on eq.~\eqref{eq:friction-theory}, a cube should diffuse more slowly than an octahedron, while an octahedron should diffuse more slowly than a tetrahedron when all have the same edge length $a$; a sphere with diameter $d = a$ is predicted to have $D_0$ between that of the octahedron and the tetrahedron (Table~\ref{table:diffusion-coeff-comparison}). Indeed, our simulation results for $D_0$ were qualitatively consistent with these predictions. Quantitatively, $D_0$ from the cube simulations was in excellent agreement with the value predicted using eq.~\eqref{eq:friction-theory}, but $D_0$ from the octahedron and tetrahedron simulations was 9\% and 18\% smaller, respectively. We calculated a similar deviation between the measured and predicted $D_0$ for tetrahedra in recent experiments by Hoffmann and coworkers, who fabricated tetrahedral clusters from four spherical polystyrene NPs with diameter $154\,\text{nm}$; they measured a self-diffusion coefficient of $D_0 = 1.72 \times 10^{-12}\, {\rm m^2/s}$ in water \cite{hoffmann:nano:2009, hoffmann:lang:2018}, which is 22\% smaller than the predicted value of $D_0 = 2.2 \times 10^{-12}\, {\rm m^2/s}$ when using an edge length of $a=308 \, {\rm nm}$ in eq.~\eqref{eq:friction-theory}. These clusters are, however, not true tetrahedra so it is unclear whether this deviation from eq.~\eqref{eq:friction-theory} should be expected in the MPCD simulations too.

\begin{table}
\caption{Diffusion coefficient at infinite dilution $D_0$ for the sphere and regular polyhedra calculated from our simulations, using eq.~\eqref{eq:friction-theory}, and using the Stokes--Einstein relationship for a sphere with mean diameter $\bar d$ given in Table~\ref{table:shape-specific-params}. All are in units of $10^{-3}\,\ell^2/\tau$.}
\label{table:diffusion-coeff-comparison}
\begin{tabular}{cccc}
     & simulation & using eq.~\eqref{eq:friction-theory}& using $\bar d$ \\ \hline
     sphere & 4.32 & 4.48 & \\
     octahedron & 3.95 & 4.36 & 4.01 \\
     cube & 3.31 & 3.33 & 3.28 \\
     tetrahedron & 5.14 & 6.29 & 5.48 \\
\end{tabular}
\end{table}

We and others previously found that $D_0$ for a cube can also be reasonably well-approximated by $D_0$ for a sphere with diameter $\bar d = (d_{\rm I} + d_{\rm C})/2$, the arithmetic mean of the diameters $d_{\rm I}$ and $d_{\rm C}$ of the spheres that inscribe and circumscribe it, respectively.\cite{wani:jcp:2022, okada:mp:2020} We carried out the same calculation for the octahedron and tetrahedron, and we again found good agreement with our simulations (Table \ref{table:diffusion-coeff-comparison}). Thus, using $\bar d$ seems to provide a quick and reasonable estimate of $D_0$ for regular polyhedra as an alternative to eq.~\eqref{eq:friction-theory}.

\begin{figure}
  \includegraphics{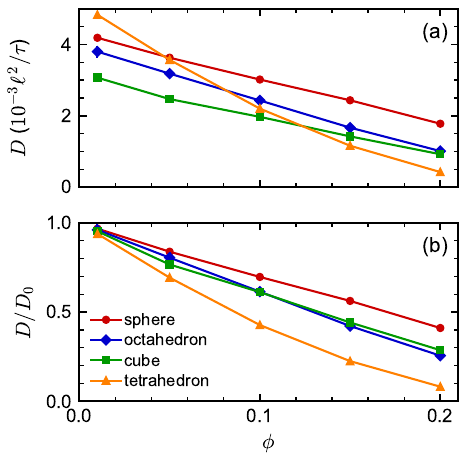}
  \caption{(a) Long-time self-diffusion coefficient $D$ of spheres, octahedra, cubes, and tetrahedra as a function of nominal volume fraction $\phi$. (b) $D$ normalized by its value linearly extrapolated to infinite dilution $D_0$.}
  \label{fig:diffusion}
\end{figure}

We next investigated the volume-fraction dependence of $D$ [Fig.~\ref{fig:diffusion}(a)]. Given that the different NP shapes had different $D_0$, we report $D/D_0$ to facilitate comparison between shapes [Fig.~\ref{fig:diffusion}(b)]. The tetrahedra exhibited the strongest dependence on $\phi$, the spheres exhibited the weakest dependence on $\phi$, while both the cubes and octahedra exhibited a similar dependence on $\phi$ that was intermediate between the spheres and tetrahedra. In general, we expected $D$ to decrease when $\phi$ increased because increased interactions between NPs usually slow their motion. At low NP volume fractions, long-ranged solvent-mediated HIs are important because short-ranged interactions are infrequent. Differences in the dependence of $D/D_0$ on $\phi$ seen in Fig.~\ref{fig:diffusion} when $\phi$ is small are then likely caused by differences in HIs between shapes.

\begin{figure}
  \includegraphics{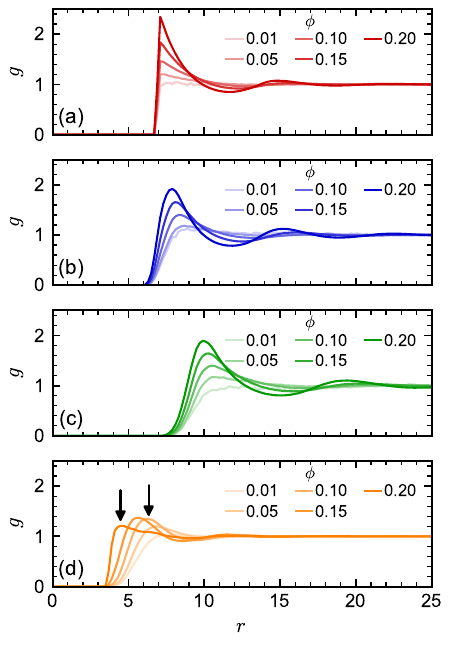}
  \caption{Radial distribution function for (a) spheres, (b) octahedra, (c) cubes, and (d) tetrahedra at different nominal volume fractions $\phi$. The arrows in (d) denote signature peaks for the transition to pentagonal dipyramids at $0.55 a_{\rm ex}$ and $0.75 a_{\rm ex}$ \cite{haji-akbari:nat:2009,haji-akbari:jcp:2011}.}
  \label{fig:rdf}
\end{figure}

At higher NP volume fractions, direct interactions between NPs become more frequent and significant, particularly those due to excluded-volume between NPs. Indeed, we expect that eventually $D/D_0 \to 0$ when the NPs reach a freezing or jamming transition that essentially traps each NP in a local cage of surrounding NPs. Based on $\phi$, the regular polyhedra we simulated were all expected to be fluids even at our largest concentration ($\phi = 0.20$).\cite{haji-akbari:nat:2009, haji-akbari:jcp:2011, agarwal:nm:2011, Gantapara:prl:2013} However, we noted that the actual excluded volume $v_{\rm ex}$ of the NPs (and hence excluded-volume fraction $\phi_{\rm ex}$) differs from the nominal volume $v$ (and nominal volume fraction $\phi$) because the vertex particles in our discrete model have a finite size. For example, the vertex particles on the surface of the cube [Fig.~\ref{fig:particles}(c)] protrude roughly $\sigma/2$, so the edge length of the volume excluded by the cube is roughly $\sigma$ longer than the nominal edge length. In general, we can estimate the excluded-volume edge length $a_{\rm ex}$ of our regular polyhedra as $a_{\rm ex} = a(1 + \sigma/d_{\rm I})$, such that the excluded-volume polyhedron contains all spheres with diameter $\sigma$ on the surface of the nominal polyhedron. The ratio of the excluded volume to nominal volume is then $v_{\rm ex}/v \approx (1+\sigma/d_{\rm I})^3$ and $\phi_{\rm ex}$ is proportionally larger than $\phi$ by the same factor. This larger excluded size $a_{\rm ex}$ is evident in the radial distribution function $g(r)$ (Fig.~\ref{fig:rdf}) for all shapes. 

\begin{figure}
  \includegraphics{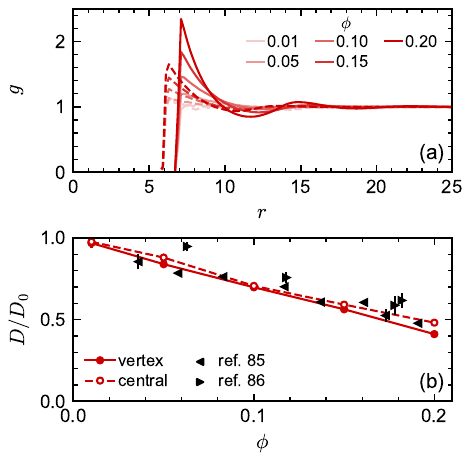}
  \caption{(a) Radial distribution functions for spheres with excluded volume handled through either vertex particles (solid lines) or a central particle (dashed lines). (b) Long-time self-diffusion coefficient $D$ (normalized by $D_0$) for the same systems compared with experimental data \cite{vanmegen:jcp:1989,vanblaaderen:jcp:1992}.}
  \label{fig:spherecompare}
\end{figure}

\begin{figure*}
  \includegraphics{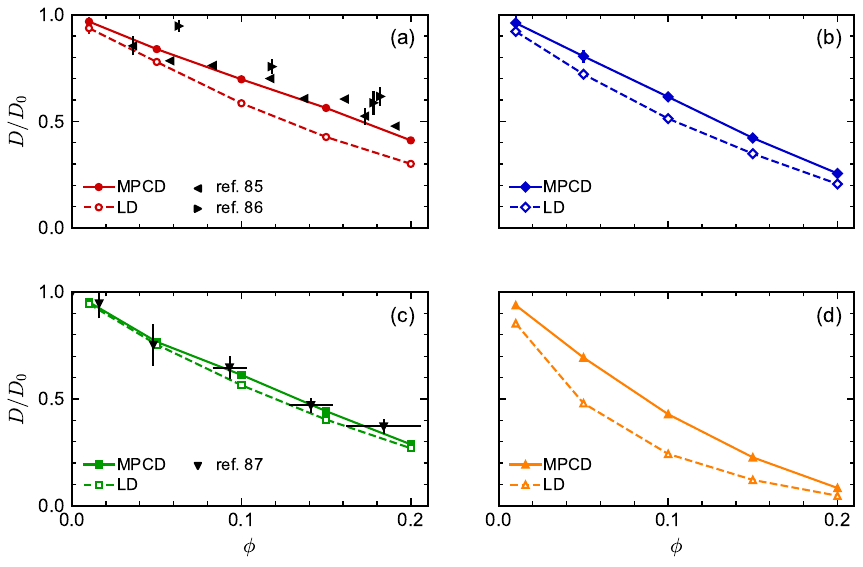}
  \caption{Comparison of long-time self-diffusion coefficient $D$ (normalized by $D_0$) of (a) spheres, (b) octahedra, (c) cubes, and (d) tetrahedra as functions of volume fraction $\phi$ from MPCD and LD simulations. Experimental data is included in (a) and (c) from multiple sources. The experimental values of $D$ for the spheres \cite{vanmegen:jcp:1989,vanblaaderen:jcp:1992} were scaled by the Stokes--Einstein prediction for $D_0$, while the experimental values of $D$ for the cubes \cite{royer:sm:2015} were scaled such that $D/D_0 = 1$ for the lowest-concentration point in that data set ($\phi \approx 0$).}
  \label{fig:diffusionld}
\end{figure*}

We attempted to assess the effect of this difference in nominal and excluded volume using the spherical NPs. We performed additional simulations where we implemented the excluded-volume interaction between spheres through a core-shifted Weeks--Chandler--Andersen potential between only their central particles, like in ref.~\citenum{wani:jcp:2022}. As expected, we found that there was less structuring in the fluid, measured through $g(r)$, at a given nominal volume fraction $\phi$ due to the smaller excluded volume of each sphere [Fig.~\ref{fig:spherecompare}(a)]. However, we found little difference in the diffusivity over the range of volume fractions investigated [Fig.~\ref{fig:spherecompare}(b)]. Moreover, the simulation data of $D/D_0$ agreed well with experimental data when plotted using $\phi$. We observed similar agreement between MPCD simulations and experiments for cubes using the nominal volume fraction $\phi$ in our prior work \cite{wani:jcp:2022} [see also Fig.~\ref{fig:diffusionld}(c)]. Hence, at least over the range of volume fractions considered for the spheres, the nominal volume fraction $\phi$ seems to be a good description of the concentration.

We note, though, that structural effects caused by differences in nominal and excluded volume may still become significant at sufficiently high excluded volume fractions, particularly if a phase transition is approached. The tetrahedron, which has the largest $v_{\rm ex}/v$ ratio of our regular polyhedra, is an excellent example of this point. Previous simulations of hard tetrahedra\cite{haji-akbari:jcp:2011, haji-akbari:nat:2009} revealed a transition to a fluid consisting of pentagonal dipyramids when the volume fraction was 0.47. That study found that $g(r)$ showed a distinct signature of this transition: at low volume fractions where dipyramids did not form, $g(r)$ had its first peak at $r = 0.75\,a$; whereas, at higher volume fractions where dipyramids formed, this original peak disappeared, and the first peak shifted to a much smaller distance $r = 0.55\,a$. Our largest nominal volume fraction $\phi = 0.20$ is well below the reported transition to dipyramids, but if we instead consider the excluded volume fraction ($\phi_{\rm ex} = 0.56$), then the system should have surpassed this transition. When we computed $g(r)$ for the tetrahedra [Fig.~\ref{fig:rdf}(d)], we observed these signature peaks emerging at the reported $r$ if $a_{\rm ex}$ was used rather than $a$. Thus, the tetrahedra appear to undergo a transition to dipyramids that is not expected using only $\phi$. The more dramatic slowing down of the tetrahedra dynamics with $\phi$ compared to the other shapes could be partially due to this transition.

Finally, we assessed the influence of HIs between the NPs on their long-time self-diffusion by performing complementary LD simulations that do not have these interactions (Fig.~\ref{fig:diffusionld}). Due to the lack of long-ranged solvent-mediated HIs, no finite-size corrections are needed. Qualitatively, $D/D_0$ had a similar dependence on $\phi$ both with and without HIs, with differences for the tetrahedra being most pronounced and differences for the cubes being least pronounced. However, there were clear quantitative differences between the MPCD simulations with HIs and the LD simulations without HIs. For all shapes, $D/D_0$ was smaller for a given $\phi$ (had a stronger $\phi$ dependence) in the LD simulations compared to the MPCD simulations. Taken together, these differences support the established picture that solvent-mediated HIs and excluded-volume interactions between NPs that determine their fluid structure both play a role in determining the NP dynamics.

As an aside, we remark that the agreement between our MPCD simulations and experiments \cite{vanmegen:jcp:1989, vanblaaderen:jcp:1992} significantly improved for the spherical NPs compared to our previous study,\cite{wani:jcp:2022} which is likely due to the higher surface density of vertex particles used in this work. The accuracy of discrete particle models typically improves with increasing surface density,\cite{swan:pof:2016} and the surface density of vertex particles on the sphere was roughly four times that of ref.~\citenum{wani:jcp:2022}. We note that Poblete et al.~recommended an optimal surface density of $0.53/\ell^2$ for spheres in MPCD to balance discretization and inertia effects,\cite{poblete:pre:2014} which lies between the value of $0.37/\ell^2$ used in ref.~\citenum{wani:jcp:2022} and $1.43/\ell^2$ used here. The surface density of vertex particles used for the other regular polyhedra was comparable to that of the spheres.

\subsubsection{Spherocylinders}
Having studied the self-diffusion of these regular polyhedra, we next investigated the long-time self-diffusion of spherocylinders. Bolhuis and Frenkel numerically studied the phase diagram of hard spherocylinders for a range of aspect ratios $\lambda$.\cite{bolhuis:jcp:1997} For $\lambda \le 2$, the spherocylinders exhibited only two phases---a low-density isotropic phase and a high-density crystal phase---with the transition between these occurring at volume fraction 0.58 and 0.53 for $\lambda = 1$ and 2, respectively. We therefore restricted our simulations to $\phi \le 0.30$, which corresponds to $\phi_\text{ex} < 0.44$ for our spherocylinders ($v_{\rm ex}/v = 1.45$ and 1.41 for $\lambda = 1$ and 2), in order to focus our calculations on the isotropic phase. We confirmed this was the case by computing a global nematic order parameter \cite{milchev:jcp:2018, yetkin:lang:2024}, finding it to be close to zero (0.02 and 0.03 for $\lambda = 1$ and 2 when $\phi = 0.30$) as expected for an isotropic phase.

In the isotropic phase, the translational diffusion of rod-like objects is the orientational average of their parallel and normal components. The self-diffusion coefficient of rods in the infinite dilution limit can be estimated as\cite{bruggen:pre:1997}
\begin{align} \label{eq:D0-rods}
    D_{0}= & \frac{k_\text{B}T}{3\pi\eta (\lambda+1)d} \bigg[ \ln(\lambda + 1)  \nonumber \\
    & + 0.316 + \frac{0.5825}{(\lambda + 1)} + \frac{0.050}{(\lambda + 1)^2} \bigg],
\end{align}
where the last three terms in the parenthesis correct for end effects \cite{tirado:jcp:1979, tirado:jcp:1980, tirado:jcp:1984}. This equation gives $D_0 = 2.94 \times 10^{-3}\,\ell^2/\tau$ and $2.41 \times 10^{-3}\,\ell^2/\tau$ for rods with $\lambda = 1$ and 2 in our MPCD solvent, respectively. Our simulated values $D_0 = 3.09 \times 10^{-3}\,\ell^2/\tau$ and $2.57 \times 10^{-3}\,\ell^2/\tau$ were within $5\%$ of eq.~\eqref{eq:D0-rods}, showing the expected  decrease of $D_0$ with $\lambda$. We also note that eq.~\eqref{eq:D0-rods} underpredicts the diffusivity of a sphere ($\lambda = 0$) by about 5\% compared to the classical Stokes--Einstein relation.

\begin{figure}
  \includegraphics{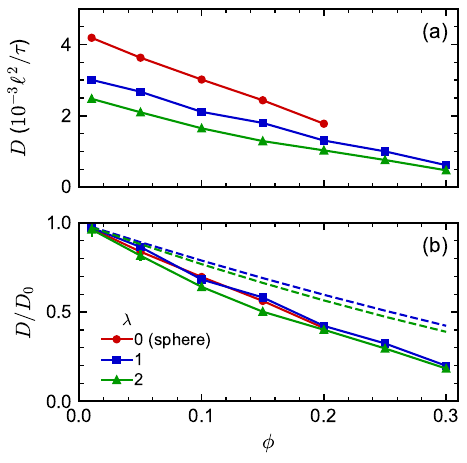}
  \caption{(a) Long-time self-diffusion coefficient $D$ for spherocylinders with aspect ratios $\lambda = 1$ and 2 as functions of volume fraction $\phi$. The sphere data from Fig.~\ref{fig:diffusion} is included as a reference point with $\lambda = 0$. (b) $D$ normalized by its value extrapolated to infinite dilution $D_0$. The dashed curves in (b) are $D/D_0$ predicted from the fit of ref.\citenum{lowen:pre:1994}.}
  \label{fig:diffusionsc}
\end{figure}

The concentration dependence of $D/D_0$ with $\phi$ was similar for both spherocylinders (Fig.~\ref{fig:diffusionsc}). Indeed, $D/D_0$ for the spherocylinders with $\lambda = 1$ was nearly indistinguishable from that for the spheres ($\lambda = 0$). The longer spherocylinders with $\lambda = 2$ showed some systematic differences, consistently having a slightly smaller value than for $\lambda = 1$ at a given $\phi$. This result indicates that even a small amount of anisotropy may begin to have an effect on the diffusive dynamics, but the magnitude of this effect seems to be small. We also compared our simulation data to the parametric fit of ref.~\citenum{lowen:pre:1994}. Our simulations qualitatively agreed with the prediction that $D/D_0$ should be smaller for a larger $\lambda$ at a given $\phi$, but the simulations consistently had smaller values of $D/D_0$ than predicted. We note that ref.~\citenum{lowen:pre:1994} used BD simulations that lacked HIs to develop this fit, so it is unclear to what extent we should expect agreement to simulations with HIs.

\subsection{Sedimentation}
\label{sec:sediment}
After investigating the long-time self-diffusion coefficients of our shape-anisotropic NPs, we characterized their sedimentation coefficients. This complementary dynamic property of a suspension is important for understanding, e.g., how NPs settle under gravity. We defined the sedimentation coefficient $K$ from the linear proportionality between the average velocity $\vv{u}$ of an NP under a sufficiently small applied force $\vv{F}$,
\begin{equation}
\vv{u} = K\gamma_0^{-1} \vv{F}.
\end{equation}
To measure $K$ in our simulations, we imposed a constant force $\vv{F} = f_x \hat{\vv{x}}$ on all NPs, where $\hat{\vv{x}}$ is the unit vector in the $x$ direction, and measured their average velocity $u_x = \vv{u} \cdot \hat{\vv{x}}$. The applied forces were $f_x = 0.5\,k_{\rm B} T/\ell$ and $1.0\,k_{\rm B} T/\ell$ per NP, which we distributed evenly among all the vertex and central particles in each NP. A balancing force was applied to the MPCD particles to ensure that the total force on the system was zero. We allowed the system to reach a steady state under the imposed forces, performed a production run where we measured $u_x$, and extracted $K$ from a linear regression of $u_x$ and $f_x$.

As for diffusion coefficients, the sedimentation coefficients from our MPCD simulations must be corrected for finite-size effects from periodic boundary conditions. The sedimentation coefficient measured in an infinitely large box $K^\infty$ is related to the one measured in a finite box by \cite{mo:pf:1994}
\begin{equation} \label{eq:K-correct}
K^\infty = K + \xi S(0) \frac{\gamma_0}{6 \pi \eta L}
\end{equation}
where $S(0)$ is the static structure factor at zero wavevector. This structure factor is related to the isothermal compressibility and so can be computed from an equation of state. Here, we used the virial expansion of the pressure, which gives
\begin{equation}
S(0) = \left(1 + \sum_{n=2} n \hat B_n \phi_{\rm ex}^{n-1} \right)^{-1}
\label{eq:s0}
\end{equation}
where $\hat B_n = B_n/v^{n-1}$ and $B_n$ is the $n$-th virial coefficient. We used $\phi_{\rm ex}$ in eq.~\eqref{eq:s0} because it should characterize the structure of the suspension better than $\phi$ (see discussion of Fig.~\ref{fig:rdf}). We used up to the 8th virial coefficient for the regular polyhedra \cite{irrgang:lang:2017} and up to the 5th virial coefficient for the spherocylinders \cite{monson:cpl:1978, barboy:jcp:1979}. Like eq.~\eqref{eq:D-FS-correct}, eq.~\eqref{eq:K-correct} also includes the suspension viscosity $\eta$ so we made the same Stokes--Einstein-like approximation to eliminate this dependency,
\begin{equation} \label{eq:K-approx}
K^\infty \approx K + \xi S(0) \frac{D}{D_0} \frac{\gamma_0}{6 \pi \eta_0 L}.
\end{equation}
We used the finite-size-corrected $D/D_0$ and computed $\gamma_0 = k_{\rm B} T/D_0$ from the finite-size-corrected $D_0$. Note that eqs.~\eqref{eq:K-correct} and \eqref{eq:K-approx} fix an error in eqs.~(19) and (20) of ref.~\citenum{wani:jcp:2022}. All sedimentation coefficients are corrected in this way, but for notational simplicity, we will refer to them as $K$ in the remaining discussion.

MPCD conserves linear momentum, so the sedimentation coefficients calculated directly from the simulation are in a frame of reference where the mass-averaged velocity of the NPs and solvent is zero. However, it is a common practice to consider suspensions in the frame of reference where the volume-averaged velocity is zero, i.e., $\langle \vv{u} \rangle = \phi \vv{u} + (1-\phi) \vv{u}_0 = \vv{0}$ where $\vv{u}_0$ is the solvent velocity. Shifting from the mass-averaged to volume-averaged frame of reference amounts to a rescaling of $K$, which we implemented as in our previous work \cite{wani:jcp:2022}. All values of $K$ are presented in the frame of reference where the volume-averaged velocity is zero.

\begin{figure}
  \includegraphics{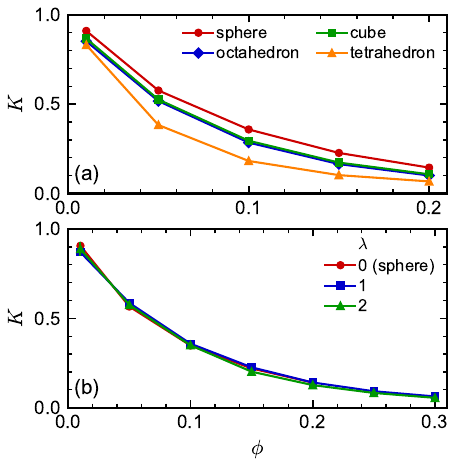}
  \caption{Sedimentation coefficient $K$ of (a) spheres, octahedra, cubes, and tetrahedra and (b) spherocylinders as a function of volume fraction $\phi$. The frame of reference used to define $K$ is the one where the volume-averaged velocity is zero.}
  \label{fig:sediment}
\end{figure}

The sedimentation coefficients of the regular polyhedra [Fig.~\ref{fig:sediment}(a)] exhibited a qualitatively similar dependence on shape and concentration as the self-diffusion coefficients did. We consistently found that the spheres had the largest $K$, the tetrahedra had the smallest $K$, while the octahedra and cubes had an intermediate $K$. Moreover, all sedimentation coefficients decreased with increasing concentration, as expected, with the tetrahedra having the strongest concentration dependence. The sedimentation coefficients of both spherocylinders were highly similar to each other and to that of the sphere [Fig.~\ref{fig:sediment}(b)]. These behaviors are qualitatively similar to the self-diffusion coefficients, so we will not repeat that discussion here for brevity.

\section{Conclusions}
We investigated the long-time self-diffusion and sedimentation of NPs with anisotropic shapes. The anisotropic shapes we studied were an octahedron, a cube, a tetrahedron, and a spherocylinder. The NPs were represented with a discrete particle model and were hydrodynamically coupled to each other using the multiparticle collision dynamics method. Simulations were conducted across a range of volume fractions for each shape where the NPs remained in a fluid/isotropic phase. Our modeling approach can be easily extended to explore the dynamics of other NP shapes, e.g., irregular polyhedra and non-convex shapes.\cite{avendano:cocis:2017}

For regular polyhedra having equal edge lengths, shape had a clear influence on their transport properties. Octahedra and cubes were slower diffusing than spheres with diameter equal to their edge length for all investigated volume fractions [Fig.~\ref{fig:diffusion}(a)]. Tetrahedra diffused the fastest at small volume fractions but the slowest at larger volume fractions, which we partially attributed to the formation of pentagonal dipyramids. The simulated self-diffusion coefficients of all investigated NP shapes at infinite dilution were in good agreement with a correlation based on sphericity and also with an approximation using the mean diameter of the spheres that inscribed and circumscribed the shapes. After accounting for differences due to shape at infinite dilution [Fig.~\ref{fig:diffusion}(b)], the self-diffusion coefficient of the spheres showed the weakest volume-fraction dependence, that of the tetrahedra showed the strongest volume-fraction dependence, while the octahedra and cubes showed intermediate behavior. Similar trends were found for the dependence of the sedimentation coefficients on volume fraction [Fig.~\ref{fig:sediment}(a)].

For small-aspect-ratio spherocylinders ($\lambda = 1$ and 2), the diffusion coefficients at infinite dilution showed a dependence on aspect ratio that was consistent with theoretical expectation, meaning that the spherocylinders diffused more slowly as aspect ratio increased [Fig.~\ref{fig:diffusionsc}(a)]. However, after accounting for shape effects at infinite dilution, the self-diffusion coefficient [Fig.~\ref{fig:diffusionsc}(b)] had a volume-fraction dependence that closely followed that of spheres having diameter equal to the spherocylinders, with only minor differences for the spherocylinder with $\lambda = 2$. The sedimentation coefficient [Fig.~\ref{fig:sediment}(b)] had essentially the same volume-fraction dependence for the spheres and both spherocylinders. We expect that the dynamics of spherocylinders should deviate more significantly from spheres as $\lambda$ increases, and in principle, we can expand our spherocylinder model to study this regime. However, doing so incurs higher computational cost due to a substantial increase in the number of vertex particles per spherocylinder. Further, we would need larger simulation boxes to accommodate these spherocylinders and gather good statistics, thereby also increasing the number of solvent particles required. To mitigate these computational challenges, an alternative approach is to represent the spherocylinders as linear rods comprised of partially overlapping particles.\cite{winkler:jpcm:2004} However, establishing a connection between this model, our spherocylinder model, and experiments is still an open question, which we plan to explore.

\section*{Author contributions}
\textbf{Yashraj M. Wani}: Conceptualization, formal analysis, investigation, methodology, visualization, writing -- original draft. \textbf{Penelope Grace Kovakas}: Conceptualization, formal analysis, investigation, methodology, writing -- review \& editing. \textbf{Arash Nikoubashman}: Conceptualization, funding acquisition, methodology, project administration, writing -- review \& editing. \textbf{Michael P. Howard}: Conceptualization, data curation, formal analysis, funding acquisition, investigation, methodology, project administration, visualization, writing -- review \& editing.

\section*{Conflicts of interest}
The authors have no conflicts to disclose.

\section*{Data Availability}
The data that support the findings of this study are available from the authors upon reasonable request.

\begin{acknowledgements}
This material is based upon work supported by the National Science Foundation under Award No.~2223084 and by the Deutsche Forschungsgemeinschaft (DFG, German Research Foundation) through Project Nos.~405552959, 470113688, and 509039598. This work was completed with resources provided by the Auburn University Easley Cluster, the supercomputer Mogon at Johannes Gutenberg University Mainz (www.hpc.uni-mainz.de), and the Texas Advanced Computing Center (TACC) at The University of Texas at Austin.
\end{acknowledgements}

\bibliography{references}

\end{document}